\documentclass[a4paper,11pt]{article}

\usepackage{jcappub} 

\title{\boldmath A search for anisotropy in the arrival directions of ultra high energy 
cosmic rays recorded at the Pierre Auger Observatory}

\author{
\par\noindent
{\bf The Pierre Auger Collaboration} \\
P.~Abreu$^{75}$, 
M.~Aglietta$^{58}$, 
M.~Ahlers$^{110}$, 
E.J.~Ahn$^{94}$, 
I.F.M.~Albuquerque$^{20}$, 
D.~Allard$^{34}$, 
I.~Allekotte$^{1}$, 
J.~Allen$^{98}$, 
P.~Allison$^{100}$, 
A.~Almela$^{13,\: 9}$
% ...

\par\noindent
$^{1}$ Centro At\'{o}mico Bariloche and Instituto 
Balseiro (CNEA-UNCuyo-CONICET), San Carlos de 
Bariloche, Argentina \\
% ...
$^{9}$ Instituto de Tecnolog\'{\i}as en Detecci\'{o}n y 
Astropart\'{\i}culas (CNEA, CONICET, UNSAM), Buenos Aires,
 Argentina \\
% ...
$^{13}$ Universidad Tecnol\'{o}gica Nacional - Facultad 
Regional Buenos Aires, Buenos Aires, Argentina \\
% ...
$^{20}$ Universidade de S\~{a}o Paulo, Instituto de 
F\'{\i}sica, S\~{a}o Paulo, SP, Brazil \\
% ...
$^{34}$ Laboratoire AstroParticule et Cosmologie 
(APC), Universit\'{e} Paris 7, CNRS-IN2P3, Paris, France 
% ...
$^{58}$ Istituto di Fisica dello Spazio 
Interplanetario (INAF), Universit\`{a} di Torino and 
Sezione INFN, Torino, Italy \\
% ...
$^{75}$ LIP and Instituto Superior T\'{e}cnico, Technical
 University of Lisbon, Portugal \\
% ...
$^{94}$ Fermilab, Batavia, IL, USA \\
% ...
$^{98}$ New York University, New York, NY, USA \\
% ...
$^{100}$ Ohio State University, Columbus, OH, USA \\
% ...
$^{110}$ University of Wisconsin, Madison, WI, USA \\
% ...
 (INST), Hanoi, Vietnam \\
\par\noindent
(‡) Deceased \\
(a) at Konan University, Kobe, Japan \\
(f) now at University of Maryland \\
(h) now at NYU Abu Dhabi \\
(i) now at Universit\'{e} de Lausanne \\
}
\emailAdd{auger.spokersperson@fnal.gov}

\abstract{
Observations of cosmic ray arrival directions made with the 
Pierre Auger Observatory have previously provided evidence of anisotropy
at the 99$\%$ CL using the correlation of ultra high energy cosmic rays (UHECRs) 
with objects drawn from the {V{\'e}ron-Cetty} {V{\'e}ron} catalog. 
In this paper we report on the use of three catalog independent methods to search 
for anisotropy.
The \mbox{2pt--L}, 2pt+ and 3pt methods, each giving a different measure
of self-clustering in arrival directions, were tested on mock
cosmic ray data sets to study the impacts of sample size and
magnetic smearing on their results, accounting for both angular
and energy resolutions.
 If the sources of UHECRs follow
the same large scale structure as ordinary galaxies in the local Universe and 
if UHECRs are deflected no more than a few degrees, a study of mock maps suggests that 
these three methods can efficiently respond to the resulting  anisotropy with a $P$-value = $1.0\%$ 
or smaller with data sets as few as 100 events. 
Using data taken from January 1, 2004 to July 31, 2010 we examined the 
20, 30, ..., 110 highest energy events with a corresponding minimum energy threshold of about
49.3 EeV.  The minimum
$P$-values found were $13.5\%$ using the 2pt-L method, $1.0\%$ using the 
2pt+ method and $1.1\%$ using the 3pt method for the highest 100 energy events.  In view of
the multiple (correlated) scans performed on the data set, these catalog-independent  methods
do not yield strong evidence of anisotropy in the highest energy cosmic rays. 
}


\begin{document}
\maketitle
\flushbottom

\section{Introduction}

It is almost 50 years since 
cosmic rays with energies of the order of 100 EeV 
(1~EeV$\equiv 10^{18}$~eV) were first reported~\cite{Linsley:1963km}. Soon after 
the initial observation of such cosmic rays it was realized by 
Greisen~\cite{Greisen:1966jv}, Zatsepin and Kuz'min~\cite{Zatsepin:1966jv} that 
their interactions  with the cosmic microwave background would result in 
energy loss that would limit the distance which they could travel. 
This would suppress the particle flux and result in a steepening of the energy
spectrum. 
If the observed flux suppression~\cite{Abbasi:2007sv,Abraham:2008ru}  is due to this
mechanism, it is likely that the cosmic rays with energies 
in excess of $\simeq 50~$EeV could be anisotropic as they would originate in the local 
Universe. Several searches for anisotropy in the arrival directions of UHECRs have been 
performed in the past, either aimed  at correlating arrival directions with astrophysical 
objects~\cite{Tinyakov:2001nr,Gorbunov:2002hk} or searching for anisotropic arrival 
directions~\cite{Stanev:1995my,Uchihori:1999gu,Tinyakov:2001ic,Takeda:1999sg}.
No positive observations have been confirmed by subsequent 
experiments~\cite{Abbasi:2008md,Abbasi:2005mk,Westerhoff:2004fj,Abbasi:2009tm}.

In 2007, the Pierre Auger Observatory~\cite{Abraham:2004dt} provided evidence 
for anisotropy at the 99\% CL (Confidence Level) by examining the correlation of UHECR ($\geq 56~$EeV) with nearby 
objects drawn from the {{V{\'e}ron-Cetty} {V{\'e}ron (VCV) catalog~\cite{VCV}. The 
correlation at a predefined 99\% CL was established with new data after 
studies of an initial  15 event-set defined a likely  increase in the UHECR flux  in circles of  $\simeq 3^\circ$ radius around Active Galactic Nuclei (AGNs) in the
VCV catalog with redshift $\leq 0.018$~\cite{Cronin:2007zz,Abraham:2007si}. An updated 
measurement of this correlation has recently been given, showing a reduced fraction of 
correlating events when compared with the first report~\cite{Abraham:2010ap}.

The determination of anisotropies in the UHECR sky distribution based on cross-correlations
with catalogs may not constitute an ideal tool in the case of large magnetic deflections
and/or transient sources. Also, some signal dilution may occur if the catalog does
not trace in a fair way the selected class of astrophysical sites, due for instance to
its incompleteness. As an alternative, we report here on tests designed to answer the question 
of whether or not the arrival directions of UHECRs observed at the Pierre Auger Observatory are 
consistent with being drawn from an isotropic distribution, with no reference to
extragalactic objects. The local Universe being distributed in-homogeneously and organized 
into clusters and superclusters, clustering of arrival directions may be expected in the case 
of relatively low source density. Hence, the methods used in this paper are based on 
searches for the self-clustering of event directions at any scale. These may thus constitute an optimal 
tool for detecting an anisotropy and meanwhile provide complementary information 
to searches for correlations between UHECR arrival directions and specific extragalactic 
objects. 

The paper is organized in the following manner. The three methods we use in this paper,
the 2pt-L, 2pt+ and 3pt methods, are explained in Section 2.  
In Section 3, we apply these methods to a toy model of anisotropy to address 
the importance of systematic uncertainties from both detector effects and unmeasured 
astrophysical parameters. In Section 4, we apply the three methods to an updated set 
of the Pierre Auger Observatory data. We draw final conclusions in Section 5.

\section{Analysis Methods}

At the highest energies, the steepening of the cosmic ray energy spectrum makes the current 
statistics so small that a measure of a statistically significant departure from isotropy 
is hard to establish, especially when using blind generic tests. This motivated us to 
develop several methods by testing their efficiency for detecting anisotropy using 
simulated samples of the Auger exposure with 60 data points drawn from different models
of anisotropies both on large and small scales. The choice of 60 events for the mock 
catalogs was based on the number of events expected for an exposure of two full years
of Auger  above the $\approx 50 $ EeV energy threshold 
for anisotropy. We report in this paper on 
self-correlation analysis, using differential approaches based on a 2pt-L
function~\cite{Peebles1980:1,Hague:2009zu}, an extended 2pt function~\cite{Ave:2009id} (``2pt+") 
and a 3pt function~\cite{Hague:2009zu} (``3pt"). 

\subsection{The 2pt-L method}
The 2-point correlation function~\cite{Peebles1980:1} is defined as 
the differential distribution over angular scale of the number of observed event pairs 
in the data set.  There are different possible implementations of statistical measures based on
the 2pt function. We adopt in this work the one in \cite{Hague:2009zu}   
(named 2pt-L in the following), where the departure from isotropy is tested through a 
pseudo-log-likelihood as described below.
The expected
distribution of the 2pt-L correlation function values was built
using a large number of simulated background sets drawn from an
isotropic
  distribution accounting for the exposure of the experiment. We use angular bins 
of 5 degrees to histogram the angle ($\lambda$) between event pairs over a range of angular scales.
The pseudo-log-likelihood  is defined as $\mathcal{L}_{2pt-L}$~:
\begin{eqnarray}
\mathcal{L}_{2pt-L}^{data}=\sum_{i=1}^{N_{bins}} \ln{P(n_{obs}^i|n_{exp}^i)},
\end{eqnarray}
where $n_{obs}^i$ and $n_{exp}^i$ are the observed and expected number of event pairs
in bin $i$ and $P$ the Poisson distribution. The resulting $\mathcal{L}_{2pt-L}^{data}$ is then
compared to the distribution of $\mathcal{L}_{2pt-L}$ obtained from isotropic Monte-Carlo samples.
The $P$-value $P_{2pt}$ for the data to come from the realization of an isotropic distribution
is finally calculated as the fraction of samples whose $\mathcal{L}_{2pt-L}$ is lower than 
$\mathcal{L}_{2pt-L}^{data}$.

\subsection{The 2pt+ method}
The 2pt-L method is sensitive to clusters of different sizes,
but not to the relative orientation of pairs. To pick up
filamentary structures or features such as excesses of pairs
aligned along some preferential directions, an enhanced version
of the classic 2pt-L test was devised: the 2pt+ 
method~\cite{Ave:2009id}. In addition
to the angular distance $\lambda$ between event pairs, the 2pt+ test
uses two extra variables related to the orientation of each vector
connecting pairs. As either one of the points in the pair can be
regarded as the vector origin, the point of origin is always chosen
so that the z-axis component is positive. This point is translated to
the center of a sphere giving rise to the two new variables, $\cos(\theta)$
which is the cosine of the vector's polar angle, and $\phi$, which
is the vector's azimuthal angle. It is worth noticing that, contrary
to the distance $\lambda$ between event pairs, these two additional
variables are not independent of the reference system in which
they are calculated. All results presented hereafter have been obtained
with the z-axis pointing toward the Northern pole in equatorial
coordinates. Fig. \ref{2pt+} schematically depicts the definitions of the variables that are used in this method. The left panel shows the vector between two points on a sphere, subtending an angle $\lambda$, which we aim to describe using three independent variables.   The vector between  two events  translated to the origin of coordinates, can be described using the following 3 variables as in the right panel of Fig. \ref{2pt+}: 
\begin{enumerate}
\item $\cos \lambda$, which is a measure of the length of the vector;
\item $\cos \theta$, which is the cosine of the vector's polar angle; and
\item $\phi$, which is the vector's azimuthal angle. 
\end{enumerate}

To measure the departure from isotropy, the 2pt-L distribution and the two angular distributions
can be combined into one single estimator.
First, in the same way as in the 2pt-L test, a binned
likelihood test is applied to the $\cos{(\lambda)}$ distribution (using a
number of bins such that the expected number of pairs in each
bin is $n_{exp}^{i}=5$), and a $P$-value
 $P_{\lambda}$ is obtained as the fraction of samples whose pseudo-log-likelihood 
$\mathcal{L}_{\lambda}$ is lower than $\mathcal{L}_{\lambda}^{data}$. Then, making use 
of the $\cos{\theta}$ and $\phi$ distributions, another pseudo-log-likelihood 
$\mathcal{L}_{\theta,\phi}$  is defined as~:
\begin{equation}
\mathcal{L}_{\theta,\phi}^{data} = \sum_{j,k=0}^{N_{bins}, \theta \phi} \ln{P(n_{j,k}^{\theta \phi, obs}, \mu)},
\end{equation} 
where $P(n_{j,k}^{\theta \phi, obs}, \mu)$ is the Poisson distribution with mean $n_{j,k}^{\theta \phi,exp}=5$ and 
$n_{j,k}^{\theta \phi, obs}$ is the observed number of pairs in the $j^{th} k^{th}$ $(\cos\theta, \phi)$ 
bin. The $P$-value $p_{\theta, \phi}$ is obtained in the same way as previously.
Finally, the combined $P$-value is calculated using Fisher's method~:
\begin{equation}
P_{combined} = P_{\lambda}p_{\theta \phi}(1- \ln P_{\lambda}p_{\theta \phi}).
\end{equation}
However, $\mathcal{L}_{\lambda}$ and $\mathcal{L}_{\theta,\phi}$ are slightly correlated tests.
Hence, $P_{combined}$ needs to be corrected for these small correlations. The final
$P$-value $P_{2pt+}$ is consequently calculated by correcting $P_{combined}$ using
Monte-Carlo simulations.

\begin{figure}[!t]
  \centering
  \includegraphics[width=0.9\textwidth]{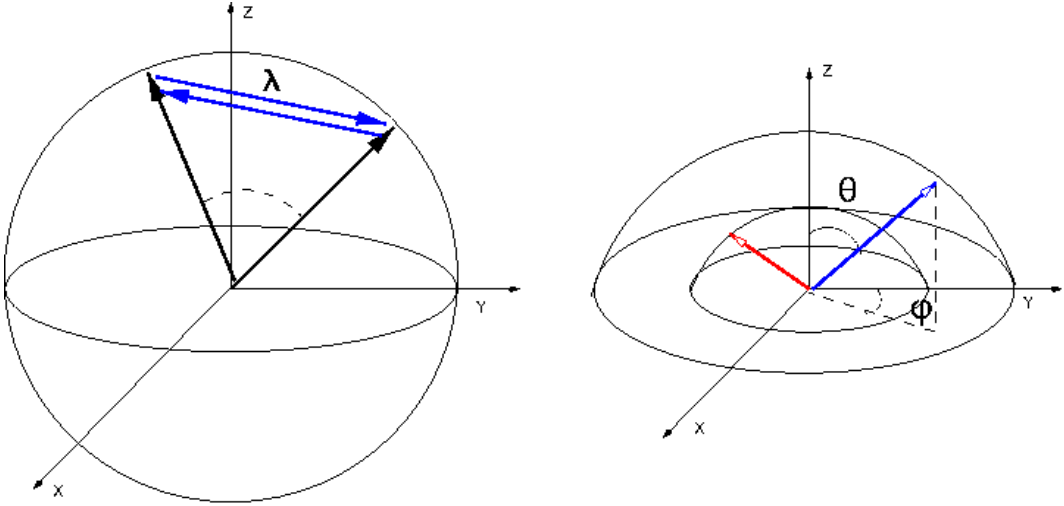}
  \caption{Schematic depiction of the variables used in the 2pt+ test. Left panel: angular distance $\lambda$ between two events. Here, the drawn sphere corresponds to the observed celestial sphere. The black vectors correspond to vectors extending from the origin (observer) to each of the two events. The blue vectors represent the two choices for the definition of the vector {\em between} events. In this work, we always use the vector with a positive $z-$component (in the case depicted here, the vector pointing from right to left). $\lambda$ is the angle subtended by the two event vectors. Right panel: angles $\theta$ and $\phi$. Here, the vectors {\em between} events have been transported to the origin, and each of the two vectors corresponds to a different pair between events. The length of each vector depends on $\lambda$, so the radii of the two spheres drawn here quantify the angular distance between each pair of events. The angles $\theta$ and $\phi$ for one of the pairs (the one represented by the blue vector) are shown in the figure. Figure from \cite{Ave:2009id}. }
   \label{2pt+}
\end{figure}

\subsection{The shape strength 3pt method}
Event direction clustering can also be revealed by
searching for excesses of triplets through, for instance, the use of a 3-pt
 correlation function.
 The 3pt method we use hereafter is
a variation of the one presented in Ref.~\cite{woodcock:1,woodcock:2}, involving an eigenvector
decomposition of the arrival directions of all triplets found in the data
set~\cite{Hague:2009zu} . For each cosmic
 ray we convert the arrival direction into a Cartesian vector 
$\vec{r}_{k}$=$\{r_{x}, r_{y}, r_{z}\}$. Then we compute an orientation matrix 
$T_{ij} = \frac{1}{3}  \sum_{k \in triplet} \left(  r_{i}r_{j} \right)_{k}$ for 
$i,j \in \{x,y,z\}$ from which we calculate eigenvalues ($\tau$) of each $T_{ij}$ 
and order them $\tau_{1}\ge\tau_{2}\ge\tau_{3}\ge{0}$ (subject to the constraint 
$\tau_{1}+\tau_{2}+\tau_{3}=1$). The largest eigenvalue of $\mathbf{T}$, $\tau_{1}$, results from a rotation of the triplet about the {\it principle} axis $\vec{u}_{1}$.
The middle and smallest eigenvalues correspond to the {\it major} $\vec{u}_{2}$ and {\it minor} $\vec{u}_{3}$ axis respectively. 
The left panel of Fig. \ref{fig:trip} shows a graphical illustration of these eigenvectors.
The eigenvalues are transformed into two parameters, 
 a ``strength parameter''$\zeta$  and  a ``shape parameter'' $\gamma$ defined as~:
\begin{eqnarray}
\zeta=\ln(\tau_{1}/\tau_{3}),
\end{eqnarray}
\begin{eqnarray}
\gamma = \ln \left\{ \frac{\ln(\tau_{1}/\tau_{2})}{\ln(\tau_{2}/\tau_{3})} \right\}.
\end{eqnarray}

As $\zeta$ increases from 0 to $\infty$ the events in the triplet become more concentrated. 
Generally, as $\gamma$ increases from $-\infty$ to $+\infty$ the shape of the triplet transforms from 
elliptical, i.e. strings, to symmetric about $\vec{u}_{1}$, i.e. point source. 
See the right panel of Fig.  \ref{fig:trip} for a schematic representation.

\begin{figure}[ht]
  \begin{tabular}{cc}
    \includegraphics[width=0.35\linewidth]{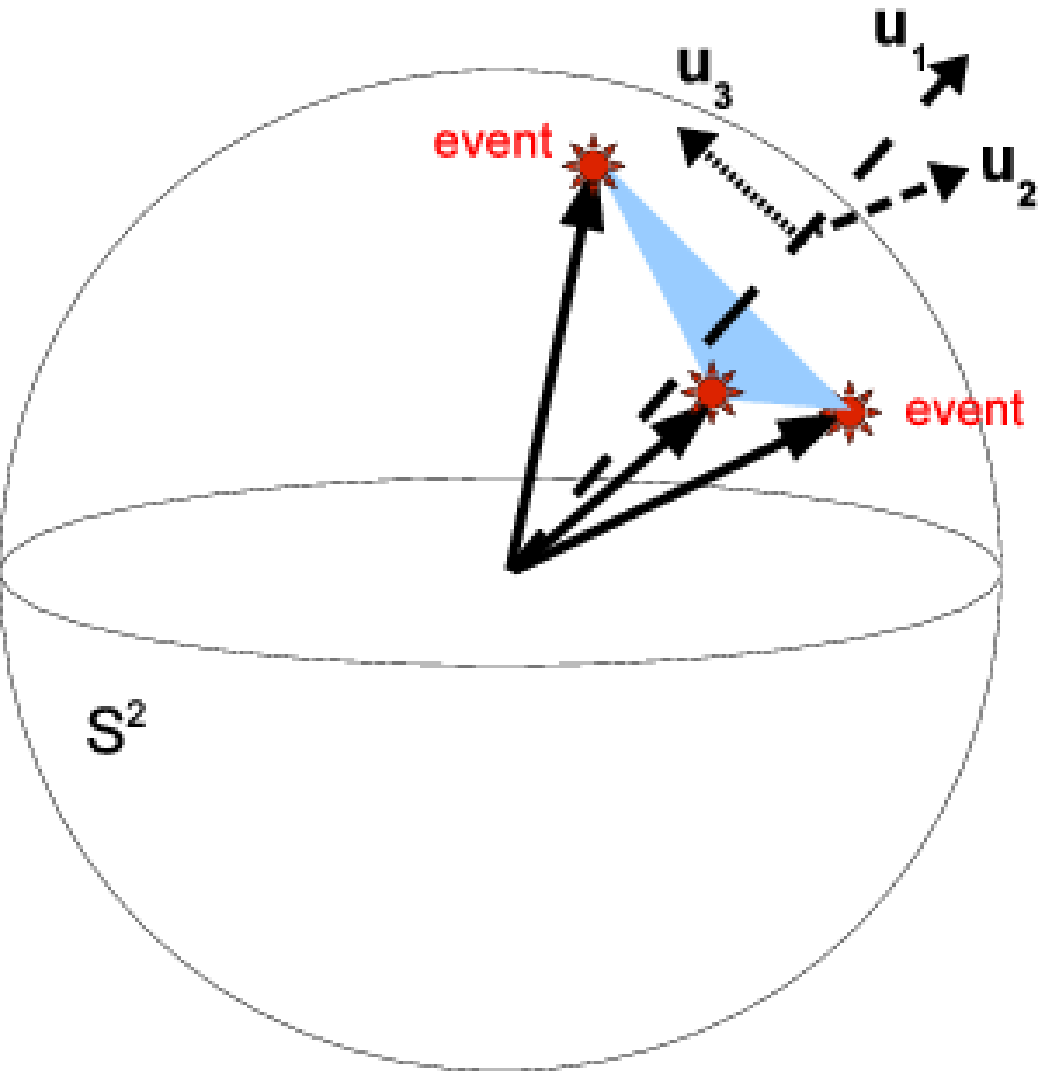} 
    \includegraphics[width=0.65\linewidth]{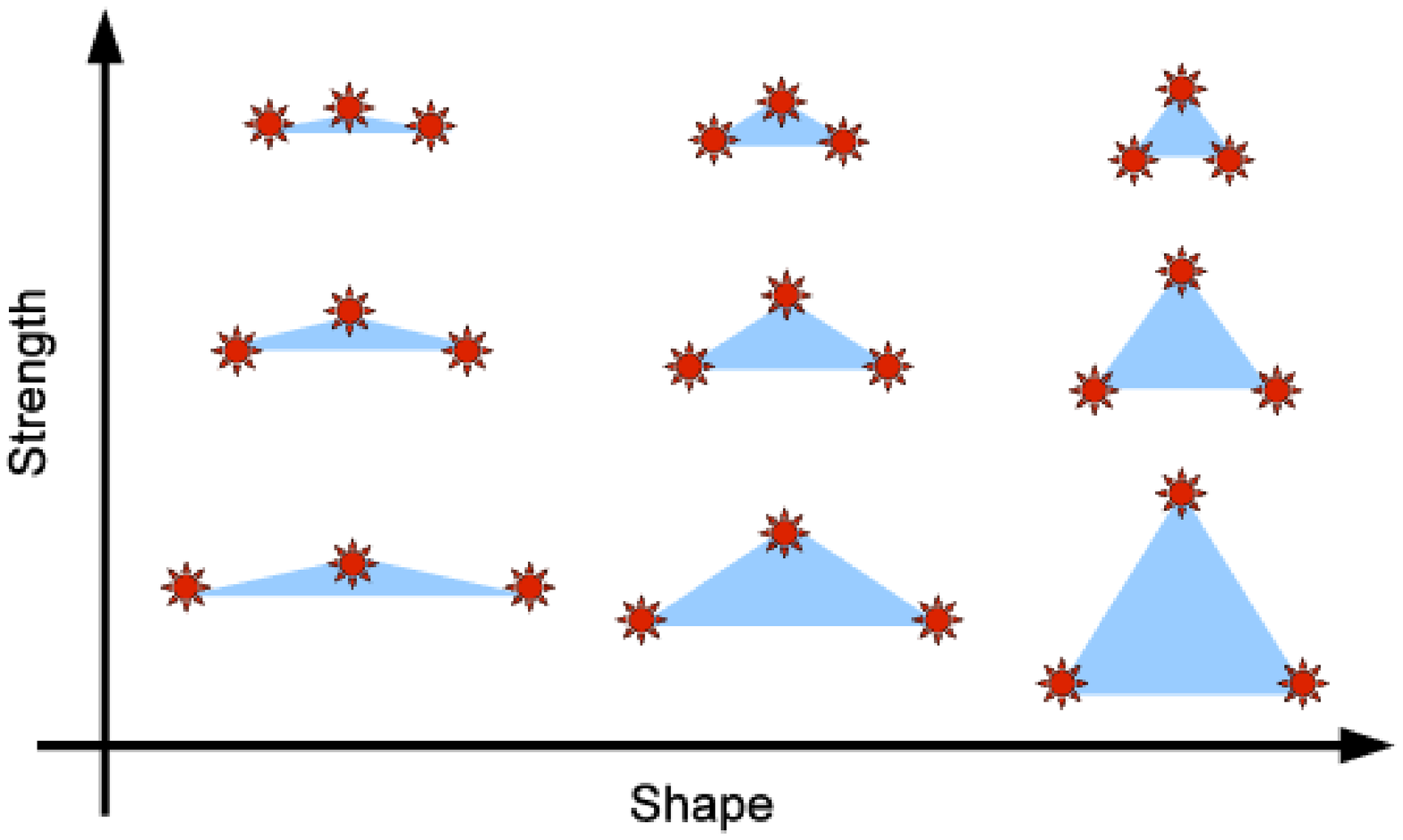} 
  \end{tabular}
   \caption{\label{fig:trip}
     {\it Left:} The eigenvectors of a triplet of events on the sphere ($S^{2}$) are the principle axis $\vec{u}_{1}$, 
     the  major axis $\vec{u}_{2}$ (pointing into the page) and the minor axis $\vec{u}_{3}$.
     The eigenvalues of these vectors are used to compute this triplet's shape and strength.
     {\it Right:} An intuitive interpretation of the shape and strength parameters.
     As the strength parameter $\zeta$ increases from $0$ to $\infty$, the events become more concentrated.
     As the shape parameter $\gamma$ increases form $-\infty$ to $+\infty$, the events become more rotationally symmetric or less elongated. Figures from \cite{Hague:2009zu}.
}
\end{figure}

After all triplets are transformed into the  parameters $\gamma$ and $\zeta$, they are 
binned in a 2-d histogram. As  $\zeta$ increases from 0 to $\infty$, the events
in the triplets become more concentrated, while, as $\gamma$ increases
from $-\infty$ to $+\infty$, the triplets transform from elongated
elliptical to symmetric shape. This distribution is then compared against the one obtained
 from all triplets on a large number of
Monte-Carlo isotropic samples.
The departure of the data from isotropy is then measured in the same way as before, 
through a pseudo-log-likelihood $\mathcal{L}_{3pt}$ where we use the Poisson distribution 
to evaluate in each bin of ($\zeta,\gamma$) the probability of observing $n_{obs}^i$ 
counts while the expected number of counts obtained from isotropic samples is $n_{exp}^i$.

\section{Application of methods to Monte-Carlo data sets}

The use of mock data sets built from the large scale structure of the Universe provides
a useful tool to study the sensitivity of the three methods. The toy model we choose
here allows us to probe the efficiency of the methods by varying several parameters
such as the total number of events, the dilution of the signal with the addition of
isotropic events, the source density, and the external smearing applied to mock data
set arrival directions. This external smearing (non angular resolution) 
reflects the unknown deflections imposed by the 
intervening galactic and extragalactic magnetic fields upon charged particles whose 
mass composition remains uncertain above $\simeq$ 40~EeV. In addition, the impact 
of both the angular and energy resolutions of the experiment can be probed in individual 
realizations of the underlying toy model.

Throughout this section, we present the performances of the three methods in terms of
the \textit{power} at different \textit{threshold} values. The \textit{threshold} $\alpha$
- or type-I error rate - is the fraction of isotropic simulations in which the null hypothesis
is wrongly rejected (i.e., the test gives evidence of anisotropy when there is no anisotropy).
 The  \textit{power}  is $1-\beta$ where $\beta$ is the type-II error rate 
which is the fraction of simulations of anisotropy in which the test result does not reject
the null hypothesis of isotropy.
 
\subsection{The toy model}

The model we chose to use is the one described in Ref.~\cite{Farrar2009a}. It
relies on (realistic large scale structure) mock-catalogs of cosmic
rays above 40~EeV,
for a pure proton composition, assuming their sources are a random
subset of ordinary 
galaxies in a simulated volume-limited survey, for various choices of
source density
which are thought to be in the relevant range~: $10^{-3.5}$, $10^{-4.0}$ and
$10^{-4.5}$~Mpc$^{-3}$. The differential spectrum at the source is
taken to be $E^{-2.3}$, and
energy losses through redshift, photo-pion production and pair production are
included. To get a realistic treatment of UHECRs in the GZK transition
region (above ~50 EeV),
a realistic volume-limited source galaxy catalog is needed to
a much larger depth than is available in present-day "all sky" galaxy
surveys. In
particular, the galaxy catalog from which the source catalog is built
must be much
denser than $10^{-3.5}$~Mpc$^{-3}$ to simulate a source catalog with
that particular
density value. Therefore,  Ref. ~\cite{Farrar2009a} made use of the "Las Damas" mock galaxy
catalogs~\cite{Farrar2009b} which were created using $\Lambda$CDM
simulations with
parameters   that are tuned to agree with Sloan Digital Sky Survey
observations.\footnote{The particular mock CR catalogs used here, along with their source galaxies, are available for downloading
from \texttt{http://cosmo.nyu.edu/mockUHECR.html}.  Analogous catalogs made subsequently, e.g., for mixed composition are also provided. }

The strength and distribution of intervening magnetic fields remain poorly known,
and large deflections may be observed even for protons~\cite{Ryu:2008}. In the absence 
of a detailed knowledge of both magnetic fields and mass composition
of CRs above 40~EeV, we smear out each arrival direction by adopting a Gaussian probability
density function with a characteristic scale ranging from 1$^\circ$ to 8$^\circ$.  
This angle is treated as a free model parameter, and each mock data set has 
a \textit{fixed} smearing angle. 

The use of a pure proton composition in this toy model
is just aimed at providing a realistic shortening of the CR horizon in the simulations
through the GZK effect. Similar behavior would be obtained in the case of heavier nuclei,
through photo-disintegration processes. The mock data sets produced with large smearing 
angles are intended to probe the lowering of the efficiencies of the methods for
situations in which  the magnetic deflections get larger, necessarily the case if the composition 
gets heavier. Smearing angles larger than 8$^\circ$ would yield almost isotropic maps,
lowering to a large extent the detection power of the methods.

Examples of sky maps produced from this toy model are shown in Fig.~\ref{SMR35},~\ref{SMR45}.
In Fig.~\ref{SMR35}, a high source density of $10^{-3.5}$~Mpc$^{-3}$ is used with an 
intermediate smearing angle of 5$^\circ$. When the 3pt method is applied to the 60 highest 
energy events from these arrival directions, a $P$-value $P_{3pt}\approx$ 0.5 is found and the 2pt+
method yelds the same value. 
On the other hand, in Fig.~\ref{SMR45}, a smaller source density is used, still with an
intermediate smearing angle of 5$^\circ$. Much more clustering can be observed. When the
3pt method is applied to the 60 highest energy events, a $P$-value $P_{3pt}\approx$ 0.003 
is found and the 2pt+ value is $P_{2pt+}\approx 0.004$.

\begin{figure}[!t]
  \centering
  \includegraphics[width=0.9\textwidth]{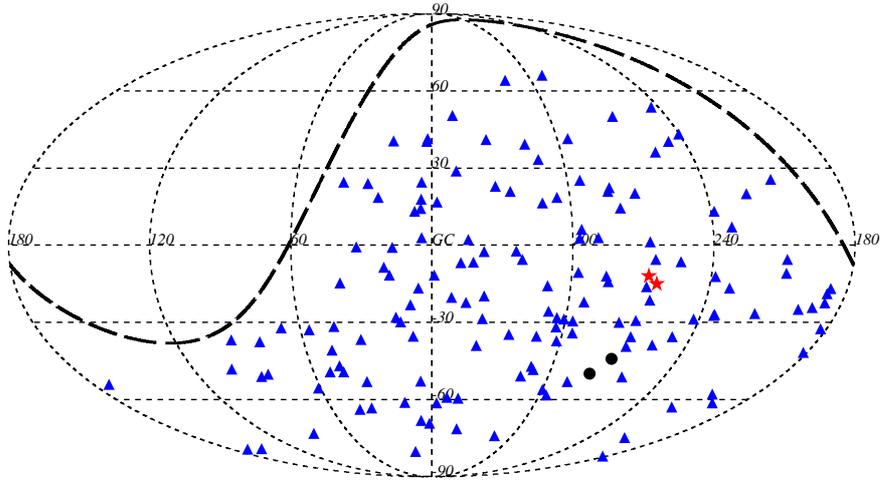}
  \caption{This map shows the Monte Carlo generated arrival directions of cosmic rays from a map with a $10^{-3.5}$~Mpc$^{-3}$ source density and a 5$^\circ$ angular smearing for 150 events. This is an example of map which is consistent with isotropy. 
  The upward triangles (in blue) are single events that come from different sources. The other symbols represent sets of events that came from the same source.
   }
   \label{SMR35}
\end{figure}
\begin{figure}[!h]
  \centering
  \includegraphics[width=0.9\textwidth]{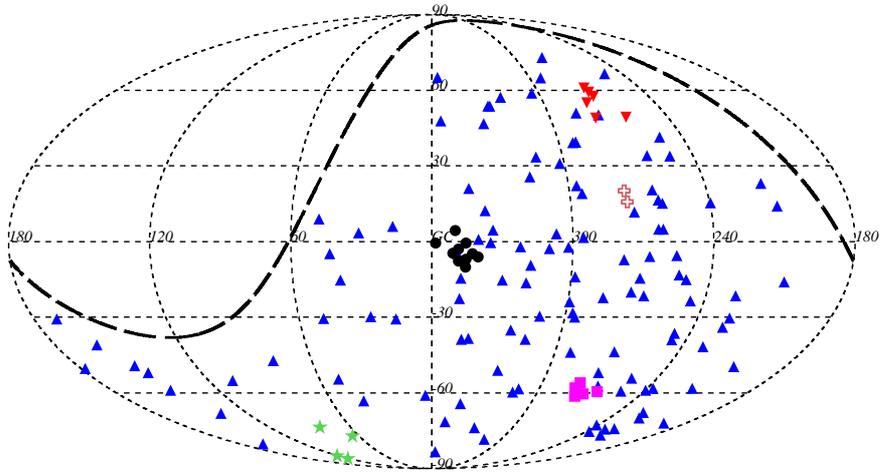}
  \caption{This map shows the Monte Carlo generated arrival directions of cosmic rays from a map with an $10^{-4.5}$~Mpc$^{-3}$ source density and a 5$^\circ$ angular smearing for 150 events. This is an example of a map which is not consistent with isotropy.
   The upward triangles (in blue) are single events that come from different sources. The other symbols represent sets of events that came from the same source.
   }
   \label{SMR45}
\end{figure}

\subsection{Application of methods to Monte-Carlo data sets}

In the toy model, the shortening of the horizon at ultra high energies (UHE) implies that CRs must come
from relatively close sources ($\lesssim$ 250~Mpc) above UHE thresholds. 
When the energy threshold is reduced, 
the CR horizon is increased and the distribution 
of sources becomes isotropic. This GZK effect induces a signal dilution as the energy
threshold is lowered, implying a loss of sensitivity of the three methods for detecting
anisotropy~\cite{Ave:2009id,Hague:2009zu}. Through this mechanism, the effects of dilution
and sample size are interconnected. We study below the efficiencies of the methods by varying
the sample size, the source density, and the external smearing. The powers of the methods 
applied to mock data sets built with a large external smearing and a high source density
of $10^{-3.5}~$Mpc$^{-3}$ are expected to be low, while an anisotropy in mock data sets built 
with a low external smearing and a low source density of $10^{-4.5}~$Mpc$^{-3}$ is expected 
to be observed with much higher powers.

\begin{figure}[!t]
  \centering
  \includegraphics[width=0.9\textwidth]{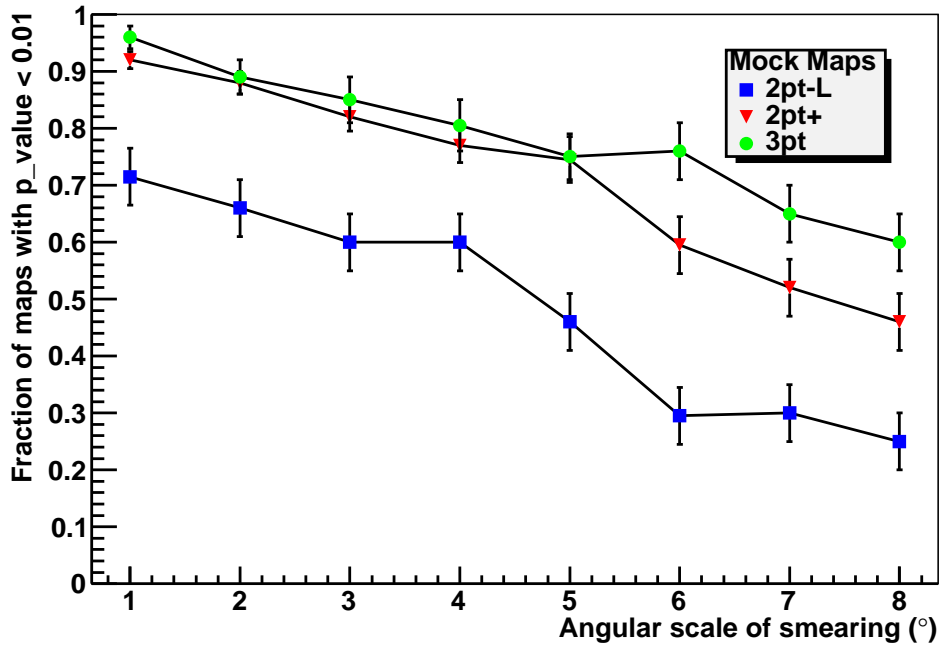}
  \caption{Powers of the 2pt-L, 2pt+ and 3pt methods for a threshold value $\alpha=1\%$. 
The mock sets shown are for 60 events drawn from maps with a source density of 
$10^{-4.5}$~Mpc$^{-3}$. The results are shown for external angular smearings ranging
from 1 to 8$^\circ$. The 2pt+ and 3pt methods perform better than the 2pt-L method 
and all three methods have decreased performances for larger angular smearing. }
   \label{powerMC1}
\end{figure}
\begin{figure}[!h]
  \centering
  \includegraphics[width=0.9\textwidth]{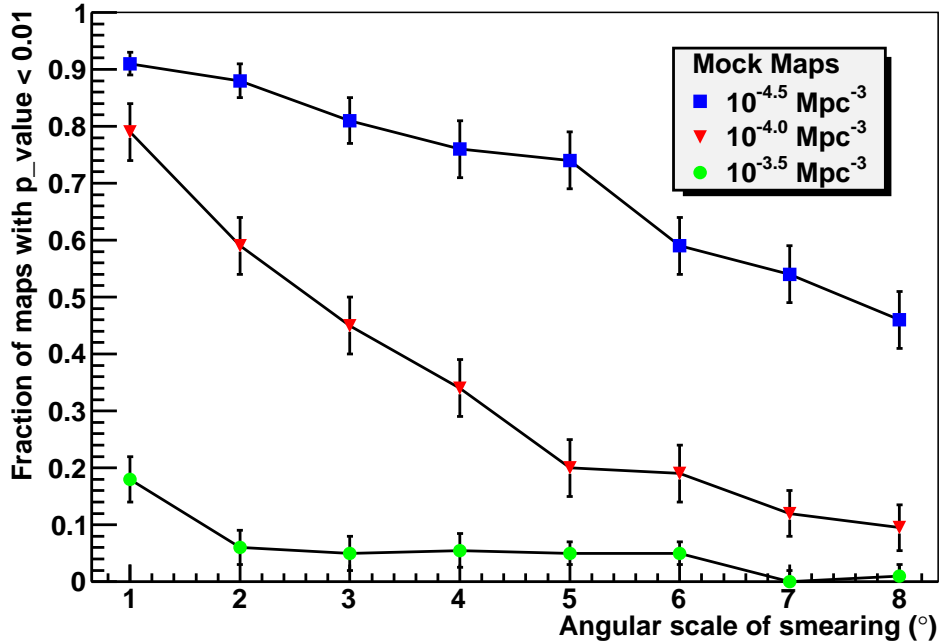}
  \caption{Power of the 2pt+ method for a threshold value $\alpha=1\%$, for different 
source density values. The mock data sets shown are for 60 events drawn from maps with source densities of $10^{-3.5}$, $10^{-4.0}$ and $10^{-4.5}$~Mpc$^{-3}$. The results 
are shown for external angular smearings ranging from 1 to 8$^\circ$. The 2pt+ method 
is clearly more efficient at smaller densities (the same behavior is also observed with 
the other two methods).}
   \label{powerMC2}
\end{figure}
\begin{figure}[!h]
  \centering
  \includegraphics[width=0.9\textwidth]{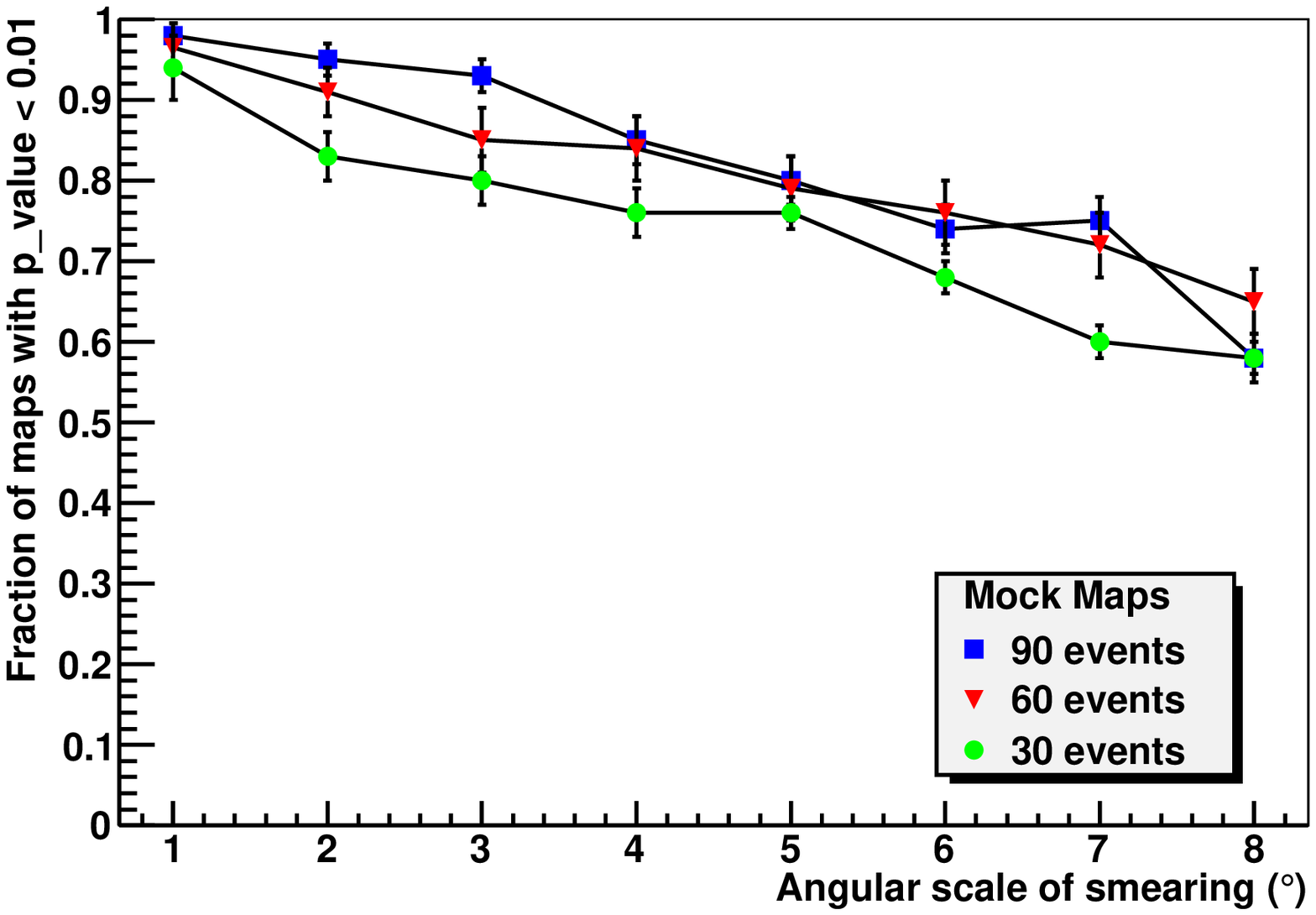}
  \caption{Power of the 3pt method for a threshold value $\alpha=1\%$, for different 
numbers of events (30, 60, and 90) and for a source density of $10^{-4.5}$~Mpc$^{-3}$. 
The results are shown for external angular smearing ranging from 1 to 8$^\circ$. The energy threshold of the highest energy 30, 60 and 90 events is equivalent to about 64, 54 and 47 EeV respectively.}
   \label{powerMC3}
\end{figure}

Finite angular and energy resolutions constitute an experimental source of signal 
dilution. Finite angular resolution is expected to slightly smooth out any clustered 
pattern, while finite energy resolution is expected to allow low energy events to leak
into higher energy populations due to the combination of the steepening of the energy
spectrum and of the sharp energy threshold used in the analysis. Above 40~EeV, the 
angular resolution\footnote{The actual angular resolution is slightly more complicated, 
as in general it is a function of energy and the number of triggered tanks. However, for events 
with energies above 40~EeV, these effects are small so that we adopt here a unique value.}, 
defined as the angular aperture $\theta_0$ around the arrival directions of CRs within 
which 68\% of the showers are reconstructed, is as good as 
$\theta_0\simeq 0.8^\circ$~\cite{Bonifazi2009}. To probe the effect of this finite 
angular resolution, the arrival direction of each event from any mock data set is smeared 
out according to the Rayleigh distribution with parameter $\theta_0/1.51$, where the 
factor 1.51 is tuned to give the previously defined angular resolution. To model 
the uncorrelated energy resolution, the energy of each mock event is smeared out according to a
Gaussian distribution centered around the original energy and with a R.M.S. $\sigma_E$ 
such that $\sigma_E/E=10\%$. This uncorrelated energy resolution value, relative to the absolute
energy scale, is a fair one, accounting for both statistical and systematic 
uncertainties at the energies E $ > $ 49 EeV reported in this paper.  ~\cite{Pesce2011}. Results shown below have been obtained by applying both 
angular and energy smearings to each mock data set.

The results of the Monte-Carlo studies are shown in Fig.~\ref{powerMC1},~\ref{powerMC2}
 and~\ref{powerMC3}. These three figures show the powers obtained with 
each method as a function of the external angular smearing ranging from 1 to 8$^\circ$.
For Fig.~\ref{powerMC1},~\ref{powerMC2}
 and~\ref{powerMC3} error bars represent the binomial uncertainties from the Monte Carlo sampling.
A general feature is the large decrease of performances for larger angular smearings.
In Fig.~\ref{powerMC1}, the power of the three methods applied to mock maps using the 
60 highest energy events drawn from a source density of $10^{-4.5}$~Mpc$^{-3}$ 
is shown. The better performances of both the 2pt+ and the 3pt methods with respect
to the 2pt-L method can be observed. In Fig.~\ref{powerMC2}, the effect of changing 
the source density from $10^{-3.5}$~Mpc$^{-3}$ to $10^{-4.0}$~Mpc$^{-3}$ and to 
$10^{-4.5}$~Mpc$^{-3}$ is shown by means of the 2pt+ method, still using the 60 highest energy
events. The 2pt+ method is clearly more efficient at smaller densities (a similar effect 
is observed for the 2pt-L and 3pt methods). In Fig.~\ref{powerMC3} the effect of using the 
30, 60 and 90 highest energy events with a low source density of $10^{-4.5}$~Mpc$^{-3}$ 
is illustrated using the 3pt method. Provided that the number of events in the sky is between 30 and 90 events, it appears that there is no strong variation in the power of the 3pt method when using such a low density. 

From these studies, it is apparent that searches for deviations from isotropic expectations 
of self-clustering at any scale, using the 2pt+ and 3pt methods, provide 
powerful tools to detect (with a threshold of 1\%) an anisotropy induced by the shortening 
of the CR horizon at UHE, even when dealing with less than 100 events. \textit{Accounting for
both angular and energy resolutions}, in the case of source densities
of the order of $10^{-4.5}$~Mpc$^{-3}$, the power of both methods is higher than 80\% as long 
as the external smearing is less than $\simeq 3-4^\circ$.  On the  other hand, for higher external smearing and/or
higher source densities, the powers rapidly decrease so that the methods may often miss 
a genuine signal in such conditions. 

\section{Application of methods to data}

\begin{table}[!ht]
\centering
\begin{tabular}{|c|c|c|c|c|}
\hline
Number of events & E-threshold (EeV) & $P$-2pt-L & $P$-2pt+ & $P$-3pt \\
\hline
\hline
20 & 75.5 & 0.113 & 0.291 & 0.271  \\
\hline
30 & 69.8 & 0.257 & 0.059 & 0.782\\
\hline
40 & 65.9 & 0.175 & 0.010 & 0.541 \\
\hline
50 & 61.8 & 0.428 & 0.045 & 0.232 \\
\hline
60 & 57.5 & 0.174 & 0.180 &  0.020 \\
\hline
70 & 55.8 & 0.455 & 0.125 & 0.154 \\
\hline
80 & 53.7 & 0.482 & 0.175 & 0.024 \\
\hline
90 & 52.3 & 0.269 & 0.013 & 0.019 \\
\hline
100 & 51.3 & 0.135 & 0.010 & 0.011 \\
\hline
110 & 49.3 & 0.239 & 0.083 & 0.075 \\
\hline
\end{tabular}
\caption{This table shows the $P$-value for the 2pt-L, 2pt+ and 3pt methods for the highest
20,30, ..., 110 energy events. The energy threshold is the energy of the lowest energy 
event in the sample. As a consequence of these sets being cumulative, the sets are correlated.
Note that none of the data sets has a $P$-value smaller than $1\%$.}
\end{table}

\begin{figure}[!t]
  \centering
  \includegraphics[width=0.9\textwidth]{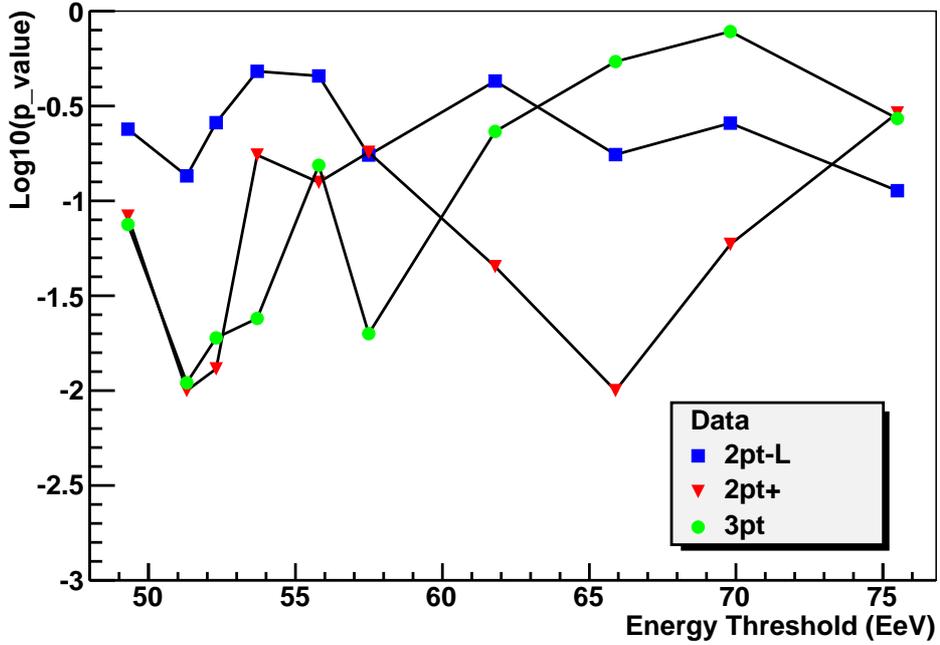}
  \caption{This shows the $P$-values of the Auger data for the 2pt-L, 2pt+ and 3pt methods. 
The minimum in $P$-value is at 100 events for the 2pt+ and 3pt methods and corresponds to 
an energy of about 51~EeV.}
   \label{fig:data}
\end{figure}

The data set analyzed here consists of events recorded by the surface detector array 
of the Pierre Auger Observatory from 1 January 2004 to 31 July 2010. During this time, 
the size of the Observatory increased from 154 to 1660 surface detector stations. In this analysis,  we 
consider  events with reconstructed zenith angles smaller 
than 60$^\circ$, satisfying fiducial cuts requiring that at least five active stations
surround the station with the highest signal, and that the reconstructed shower core
is inside a triangle of active detectors when the event was recorded. At UHE, these
requirements ensure both a good quality of event reconstruction and a robust estimation 
of the exposure of the surface detector array, which amounts to 23,344~km$^2$~sr~yr for 
the time period used in this analysis. This exposure is 2.6 times larger than that used  in Ref.~\cite{Cronin:2007zz}. 

The results of the three methods applied to the data from the 20 highest
energy events to the highest 110 are presented in Tab.~1 and shown in Fig.~\ref{fig:data}. 
The strongest deviation from isotropic expectations is found at 100 events, corresponding
to an energy threshold of $\simeq$ 51~EeV. The minimum $P$-values  are $13.5\%$ using 
the 2pt-L method, $1.0\%$ using the 2pt+ method, and $1.1\%$ using the 3pt
method. In view of the multiple scans performed, these tests do not provide strong evidence 
of anisotropy.

In case there is a true weak  anisotropy in the data, the common minimum reached by the  two more powerful methods (3pt and 2pt+)  around $E_{threshold} \approx 51 $ EeV could  indicate  the onset of this anisotropy,  while the less powerful method (2pt-L) would not have detected it. For higher energy thresholds the number of events decrease and the power of the methods diminish as expected.   If there is no weak anisotropy in the data, the Fig.~\ref{fig:data} shows only random values of $ P_{value} $ for all three methods and the common minimum mentioned before would be only a coincidence.

In our recent update on the correlation within 3.1$^\circ$ of UHECRs ($\geq 56~$EeV) with 
nearby objects drawn from the {{V{\'e}ron-Cetty} {V{\'e}ron (VCV) catalog, we reported 
a correlating fraction of $(38^{+7}_{-6})$\%, compared to 21\% for isotropic
cosmic rays. It is worth examining whether the null result reported here is compatible
with this correlating fraction or not. For this purpose, we generated mock data sets
of 80 events drawn by imposing a correlating fraction of 38\% and applied both the
2pt+ and the 3pt tests on each mock data set. The detection power of the 2pt+ (3pt) test
was found to be 10\% (20\%). These are rather low efficiencies, so that results
of the correlating fraction approach and the one chosen in this study are found to be
compatible.

\section{Conclusion}

In this paper, we have searched for self-clustering in the arrival directions of UHECRs detected at the Pierre Auger
Observatory, independently of any astrophysical catalog of extragalactic objects and
magnetic field hypothesis. These methods have been shown, within some range of parameters 
such as the magnetic deflections and the source density, to be sensitive to anisotropy in data sets drawn from mock maps which account for clustering from the large scale structure 
of the local Universe and for energy loss from the GZK effect. When applied to the highest 
energy 20, 30, ..., 110 Auger events, it is found that for the 100 highest energy events, corresponding
to an energy threshold of $\simeq$ 51~EeV, the $P$-values of 2pt+ and 3pt methods are 
about $1\%$. There is no $P$-value smaller than $1\%$ in any of the 30 (correlated) scanned 
values. There is thus no strong evidence of clustering in the data set which was examined. 

Despite of the sensitivity improvement that the 2pt+ and 3pt tests bring with respect
to the 2pt-L test, they still show relatively low powers in the case of large magnetic
deflections and/or relatively high source density. In such low event number scenarios, the search for
self-clustering of UHECRs is most likely not the optimal tool to establish anisotropy 
using the blind generic tests we presented in this paper.

\acknowledgments
 The successful installation, commissioning, and operation of the Pierre Auger Observatory
would not have been possible without the strong commitment and effort
from the technical and administrative staff in Malarg\"ue.

We are very grateful to the following agencies and organizations for financial support: 
Comisi\'on Nacional de Energ\'ia At\'omica, 
Fundaci\'on Antorchas,
Gobierno De La Provincia de Mendoza, 
Municipalidad de Malarg\"ue,
NDM Holdings and Valle Las Le\~nas, in gratitude for their continuing
cooperation over land access, Argentina; 
the Australian Research Council;
Conselho Nacional de Desenvolvimento Cient\'ifico e Tecnol\'ogico (CNPq),
Financiadora de Estudos e Projetos (FINEP),
Funda\c{c}\~ao de Amparo \`a Pesquisa do Estado de Rio de Janeiro (FAPERJ),
Funda\c{c}\~ao de Amparo \`a Pesquisa do Estado de S\~ao Paulo (FAPESP),
Minist\'erio de Ci\^{e}ncia e Tecnologia (MCT), Brazil;
AVCR AV0Z10100502 and AV0Z10100522,
GAAV KJB100100904,
MSMT-CR LA08016, LC527, 1M06002, MEB111003, and MSM0021620859, Czech Republic;
Centre de Calcul IN2P3/CNRS, 
Centre National de la Recherche Scientifique (CNRS),
Conseil R\'egional Ile-de-France,
D\'epartement  Physique Nucl\'eaire et Corpusculaire (PNC-IN2P3/CNRS),
D\'epartement Sciences de l'Univers (SDU-INSU/CNRS), France;
Bundesministerium f\"ur Bildung und Forschung (BMBF),
Deutsche Forschungsgemeinschaft (DFG),
Finanzministerium Baden-W\"urttemberg,
Helmholtz-Gemeinschaft Deutscher Forschungszentren (HGF),
Ministerium f\"ur Wissenschaft und Forschung, Nordrhein-Westfalen,
Ministerium f\"ur Wissenschaft, Forschung und Kunst, Baden-W\"urttemberg, Germany; 
Istituto Nazionale di Fisica Nucleare (INFN),
Ministero dell'Istruzione, dell'Universit\`a e della Ricerca (MIUR), Italy;
Consejo Nacional de Ciencia y Tecnolog\'ia (CONACYT), Mexico;
Ministerie van Onderwijs, Cultuur en Wetenschap,
Nederlandse Organisatie voor Wetenschappelijk Onderzoek (NWO),
Stichting voor Fundamenteel Onderzoek der Materie (FOM), Netherlands;
Ministry of Science and Higher Education,
Grant Nos. N N202 200239 and N N202 207238, Poland;
Funda\c{c}\~ao para a Ci\^{e}ncia e a Tecnologia, Portugal;
Ministry for Higher Education, Science, and Technology,
Slovenian Research Agency, Slovenia;
Comunidad de Madrid, 
Consejer\'ia de Educaci\'on de la Comunidad de Castilla La Mancha, 
FEDER funds, 
Ministerio de Ciencia e Innovaci\'on and Consolider-Ingenio 2010 (CPAN),
Xunta de Galicia, Spain;
Science and Technology Facilities Council, United Kingdom;
Department of Energy, Contract Nos. DE-AC02-07CH11359, DE-FR02-04ER41300,
National Science Foundation, Grant No. 0450696,
The Grainger Foundation USA; 
ALFA-EC / HELEN,
European Union 6th Framework Program,
Grant No. MEIF-CT-2005-025057, 
European Union 7th Framework Program, Grant No. PIEF-GA-2008-220240,
and UNESCO.


\end{document}